\documentclass[aps,prb,superscriptaddress,twocolumn,floatfix]{revtex4-1}

\usepackage{graphicx,graphics}
\usepackage{dcolumn}
\usepackage{bm}
\usepackage{braket}
\usepackage{nicefrac}
\usepackage{feynmp}
\usepackage{siunitx}
\usepackage[utf8]{inputenc}
\usepackage[abs]{overpic}
\usepackage{dcolumn}
\usepackage{amsmath,amssymb,amsfonts}
\usepackage{wasysym}
\usepackage{latexsym,verbatim}
\usepackage{color}
\usepackage[framemethod=default]{mdframed}
\usepackage[breaklinks=true,colorlinks,citecolor=blue,linkcolor=blue,urlcolor=blue]{hyperref}
\usepackage[makeroom]{cancel}
\usepackage{fancybox}

\usepackage{natbib}

\usepackage{ulem}

\usepackage[abs]{overpic}

\begin{document}

\title{Corbino FETs in magnetic field: highly tunable photodetectors}
\author{Bailey Winstanley}
\email{bailey.winstanley@manchester.ac.uk}
\affiliation{\noindent Department of Physics and Astronomy, University of Manchester, Manchester M13 9PL, UK}%
\author{Henning Schomerus}
\affiliation{\noindent  Department of Physics, Lancaster University, Lancaster, LA1 4YB, UK}%
\author{Alessandro Principi}
\email{alessandro.principi@manchester.ac.uk}
\affiliation{\noindent Department of Physics and Astronomy, University of Manchester, Manchester M13 9PL, UK}%


\begin{abstract}
We study gated field effect transistors (FETs) with an eccentric Corbino-disk geometry, such that the drain spans its circumference while the off-center inner ring acts as a source. An AC THz potential difference is applied between source and gate while a static source-drain voltage, rectified by the nonlinearities of FET electrons, is measured. When a magnetic field is applied perpendicular to the device, a strong resonance appears at the cyclotron frequency. The strength of the resonance can be tuned by changing the eccentricity of the disk. We show that there is an optimum value of the eccentricity that maximizes the responsivity of the FET.
\end{abstract}

\maketitle


\section{Introduction}

Electromagnetic radiation is one of the prime tools to investigate matter and its properties. This is made possible by the existence of efficient and compact sources and detectors in the whole spectrum, with the crucial exception of the low-THz range (between $0.1$ and $30~{\rm THz}$). This fact, commonly referred to as the ``terahertz gap'', has slowed down technological developments in, e.g., nondestructive imaging, biosensing, and spectroscopy of materials [\onlinecite{Kleiner,pablo,deng}]. In modern optoelectronics there is a deep need for efficient and tunable photodetectors that operate in this range [\onlinecite{Kleiner,pablo,deng,grigorenko,bonaccorso,tonouchi}]. Dyakonov and Shur, in 1996, predicted that a field effect transistor (FET), or any gated two-dimensional (2D) electron liquid, could be used to generate and detect THz radiation [\onlinecite{DS1,DS2,DS3}]. 

The device in their seminal work consists of a square semiconductor quantum-well cavity, hosting a 2D electron gas, connected to a source and a drain and in close proximity to a metal gate. When a THz AC source-gate voltage is applied, typically from incoming THz radiation impinging on an antenna, asymmetric boundary conditions and intrinsic nonlinearities of the electron fluid produce a rectified DC source-drain voltage. Resonances are observed in the rectified (photo)voltage at frequencies that allow plasmons (collective long-wavelength charge density fluctuations [\onlinecite{giuliani}]) to undergo constructive interference. This phenomenon has been experimentally verified in semiconductor quantum wells at room [\onlinecite{ex1,ex2,ex3}] and low temperatures [\onlinecite{knap1}] and in graphene-based FETs [\onlinecite{tomadin,yan,Vicarelli1,principi,koppens}].

Recently, it has been shown that the responsivity of Dyakonov-Shur THz detectors can be greatly enhanced by shaping them as Corbino disks [\onlinecite{bandurin-corbino}]. In such geometry, the electric field becomes singular at the inner contact ring (the source), and the field enhancement results in a strong nonlinear rectification at the outer ring (the drain). Motivated by such findings, here we study similar photodetectors in a uniform magnetic field perpendicular to the electron liquid, previously performed in other geometries and shown to enhance photodetection [\onlinecite{boubanga-tombet,bialek}]. Under this condition, the spectrum of plasmon modes, labelled by their ``winding number'' $\eta$, {\it i.e.} the number of oscillations of the electric field in the angular direction, is recontructed. Notably, the plasmon spectrum splits into two parts, revealing both bulk and edge modes. Edge magnetoplasmons have frequencies below the cyclotron frequency for values of $\eta$ that are not too large. Bulk-plasmons' frequencies are instead ``pushed'' above the cyclotron frequency. 

As shown in what follows, the energy of magnetoplasmons depends on the sign of $\eta$, with edge modes appearing only at positive winding numbers (for magnetic fields along the direction orthogonal to the disk). Furthermore, depending on device parameters and at odds with Corbino disks characterized by symmetric boundary conditions [\onlinecite{glattli,fetter2,reboredo}], the dispersion of bulk modes can exhibit a nearly-flat band close to the cyclotron frequency. When the radii of the source and drain rings are comparable, modes characterized by different winding numbers appear to have all very similar frequencies. Because of this feature, we would expect the response of the system to be greatly enhanced when the frequency of the external field is close to the cyclotron one, if we would be able to excite plasmon modes with different winding numbers at once. Since the cyclotron frequency can be tuned with the external magnetic field, the Corbino photodetector could be capable of selectively detecting frequencies deep in the THz gap with a high responsivity. 
Unfortunately, in the Corbino geometry this would require a careful fine-tuning of the potential profile at the source (inner) ring, which is highly unlikely to be realized experimentally with a simple circular contact connected to an antenna. The circular symmetry of the Corbino disk indeed forbids the mixing of modes of different winding numbers, and therefore a homogeneous potential at the source would only excite non-winding plasmons with $\eta=0$.

To overcome this limitation, we study an ``eccentric'' Corbino geometry, whereby the inner source ring is off-centered and made closer to the outer edge on one side of the disk. By breaking the circular symmetry, the eccentric geometry enables the excitation of modes characterized by different winding numbers with a simple uniform source potential. The photoresponse is greatly enhanced at frequencies near the cyclotron one when the source is in close proximity of the drain. This requirement is reminiscent of the condition needed to obtain a plasmon flat band in concentric Corbino geometries.
Therefore, in eccentric geometries, the photoresponse enhancement is controlled not only by the size of the inner ring, but also by its closeness to the drain. We find that, for any pair of source and drain radii, there is an {\it optimal} value of the eccentricity that maximizes the photoresponse.

In Sect.~\ref{sect:cavity_model} we present the model of the electron cavity as a hydrodynamic fluid in the presence of a uniform perpendicular magnetic field. 
In Sect.~\ref{sect:corbino_disk_analytic} we apply said theory to model a Corbino disk. 
In Sect.~\ref{sect:eccentric_corbino_disk} we study an eccentric Corbino disk. In Sect.~\ref{sect:conclusions} we report the summary of our findings and our main conclusions. We note that the description we use holds for a variety of different systems~[\onlinecite{DS1,DS2,DS3,principi,bandurin}], and therefore our predictions have a broad range of applicability.

\section{The model of the cavity}
\label{sect:cavity_model}

We consider a general FET, where the active component is a 2D electron liquid placed in close proximity to a metal gate. The geometry used in this paper is that of a Corbino disk with source and drain electrodes attached to the inner and outer edges, respectively. It should be noted, however, that the following applies to general 2D geometries. A radiation field oscillating at frequency $\omega$ is applied between the source and the gate, typically via an antenna, while the drain is left fluctuating, {\it i.e.} no current flows through it. We will study rectification of the oscillating field due to the intrinsic hydrodynamic nonlinearities of the electron liquid~[\onlinecite{principi,Sun,Rostami,Mikhailov1,Mikhailov2,Sipe1,Sipe2}] (we discuss below the applicability of such model).  A rectified DC source-drain potential difference, proportional to the power of the incident radiation, is therefore measured between source and drain at zero applied bias.

Since we focus on the long-wavelength low-frequency dynamics of the electron liquid, we model it by means of hydrodynamic equations~[\onlinecite{principi,principi4}]. These govern the relationship between the density, current and electric field within the device. We stress that equations formally equivalent to hydrodynamic ones can be derived by inverting the nonlinear relation between current and electric field of the electron fluid~[\onlinecite{principi}], with no reference to typical scattering times~[\onlinecite{principi1}] ({\it i.e.} the relations hold true also for non-interacting electrons). Therefore hydrodynamic equations should be seen here as an efficient way to incorporate nonlinearities in the long-wavelength description of the electron liquid. 
The first of these relations is the continuity equation,
$\partial_t \rho({\bm r},t)+\nabla\cdot[\rho({\bm r},t){\bm v}({\bm r},t)] = 0$,
which connects the periodic accumulation of charge density due to the oscillating radiation field, $\rho({\bm r},t)$, to the flow velocity, ${\bm v}({\bm r},t)$. Since electrons are charged, $\rho({\bm r},t)$
induces a nonlocal Hartree-like electric potential according to~[\onlinecite{principi}]
\begin{equation} \label{eq:U_n_def}
U({\bm r},t) = \int d{\bm r}' V({\bm r}-{\bm r}') \rho({\bm r}',t)
~,
\end{equation}
which in turn acts as the restoring force that sustains charge oscillations in a feedback loop. In Eq.~(\ref{eq:U_n_def}), $V({\bm r}-{\bm r}')$ is the Coulomb interaction between two charges at positions ${\bm r}$ and ${\bm r}'$. The nearby gate, which we assume to be a perfect conductor, has an important effect: mirror charges screen the tail of the Coulomb interaction and make it effectively short-ranged. In view of this fact, and to simplify the following derivation, we will employ the so-called ``local-gate approximation''~[\onlinecite{tomadin,principi}]. The latter consists in assuming a local relation between the self-induced field and charge density,
\begin{equation} \label{eq:local_gate_approx}
U({\bm r},t)=\rho({\bm r},t)/C
~,
\end{equation}
in lieu of the nonlocal one of Eq.~(\ref{eq:U_n_def}). This approximation has been shown~[\onlinecite{fetter1,fetter2}] to well reproduce results obtained with Eq.~(\ref{eq:U_n_def}) when the gate is explicitly accounted for. In the specific case under consideration, it allows for the emergence of edge magnetoplasmons in both semi-infinite planes and hollow disks. Using the local-gate relation between electric potential and charge density, the continuity equation becomes
\begin{equation} \label{eq:hydrodynamic1}
    \partial_t U({\bm r},t)=-\nabla\cdot[U({\bm r},t){\bm v}({\bm r},t)]
    ~.
\end{equation}
The equation relating the flow velocity to the self-induced field is assumed to have the following Euler-like form~[\onlinecite{principi}]
\begin{eqnarray} \label{eq:hydrodynamic2}
&&
\frac{e}{m}\nabla U({\bm r},t) = \partial_t{\bm v}({\bm r},t)+ \frac{1}{\tau}{\bm v}({\bm r},t)+\omega_c \hat{{\bm z}}\times{\bm v}({\bm r},t)
\nonumber\\
&&
+
\big[ {\bm v}({\bm r},t)\cdot \nabla] {\bm v}({\bm r},t) + {\bm v}({\bm r},t)\times\big[\nabla \times {\bm v}({\bm r},t)\big]
~.
\end{eqnarray}
In these equations, $-e$ is the electron charge, $m$ their effective mass and $\tau$ the average time between two successive momentum-non-conserving collisions with impurities or phonons. Finally,
$\omega_c=eB/m$ is the cyclotron frequency and $B$ is the magnetic field applied orthogonal to the 2D electron liquid. The term ${\bm v}({\bm r},t)\times\big[\nabla \times {\bm v}({\bm r},t)\big]$, known as the Lamb vector, represents a nonlinear Lorentz force due to the vortical movement of the electron fluid itself~[\onlinecite{fluid}], and can be combined with the term $\big[ {\bm v}({\bm r},t)\cdot \nabla] {\bm v}({\bm r},t)$ into the single term $\nabla v^2({\bm r},t)/2$.
We solve the problem posed by the hydrodynamic equations~(\ref{eq:hydrodynamic1})-(\ref{eq:hydrodynamic2}) in conjunction with the usual Dyakonov-Shur boundary conditions
\begin{equation}
    \begin{aligned}
    &U({\bm r},t)|_{\rm source}= U_{\rm ext}({\bm r}) \cos(\omega t)~,\\
    &{\hat {\bm n}}\cdot{\bm v}({\bm r},t)|_{\rm drain}=0~,
    \label{eq:boundary1}
    \end{aligned}
\end{equation}
corresponding to an oscillating gate-source potential from the antenna output and an open-circuit drain. Here ${\hat {\bm n}}$ is the unit vector normal to the drain surface. 

To solve the problem above, we resort to a perturbative treatment of the system of nonlinear equations. We assume $U_{\rm ext}$ to be a small parameter and calculate the rectified nonlinear response as a perturbation to the potential. We then expand
\begin{equation}
    \begin{aligned}
    &U({\bm r},t)=U_0+ U_1({\bm r},t)+ U_2({\bm r},t)+\mathcal{O}(U_{\rm ext}^3),\\
    &{\bm v}({\bm r},t)= {\bm v}_1({\bm r},t) + {\bm v}_2({\bm r},t)+\mathcal{O}(U_{\rm ext}^3) .
    \label{eq:expansion1}
    \end{aligned}
\end{equation}
Here $U_0<0$ is the equilibrium  gate potential (which fixes the charge density in the FET according to $\rho_0 = C U_0$), and the equilibrium velocity, ${\bm v}_0$, is zero by definition. $U_1({\bm r},t)$ and ${\bm v}_1({\bm r},t)$, and $U_2({\bm r},t)$ and ${\bm v}_2({\bm r},t)$ are the linear (order $U_{\rm ext}$) and nonlinear (order $U_{\rm ext}^2$) contributions to the potential and velocity, respectively. Note that, although small, $U_2({\bm r},t)$ is responsible for the only nontrivial DC rectified potential, which can be detected by measuring an averaged source-to-drain voltage drop~[\onlinecite{DS1,DS2,DS3,tomadin,principi}].

Plugging the expansions in Eq.~(\ref{eq:expansion1}) into the set of equations~(\ref{eq:local_gate_approx})-(\ref{eq:boundary1}), we collect terms of order $U_{\rm ext}$ and $U_{\rm ext}^2$ into two systems of differential equations, which are linear in $U_1({\bm r},t)$ and ${\bm v}_1({\bm r},t)$, and $U_2({\bm r},t)$ and ${\bm v}_2({\bm r},t)$, respectively. The former yields the linear response of the system which oscillates at the same frequency as the external source-gate perturbation potential, {\it i.e.}  $U_1({\bm r},t)=U_1({\bm r})e^{-i\omega t}+U_1^*({\bm r})e^{i\omega t}$ and ${\bm v}_1({\bm r},t)={\bm v}_1({\bm r})e^{-i\omega t}+{\bm v}_1^*({\bm r})e^{i\omega t}$.
Conversely, the system of equations for $U_2({\bm r},t)$ and ${\bm v}_2({\bm r},t)$ yields solutions oscillating at $\pm 2 \omega$ and a rectified (time-independent) one. To focus on the latter part of the potential $U_2({\bm r},t)$, we average equations over time by integrating over a period of oscillation, $T=2\pi/\omega$. In this way, the time-dependent parts of $U_2({\bm r},t)$ and ${\bm v}_2({\bm r},t)$ vanish. 

The details of the derivation are given in App.~\ref{app:derivation_model}. The linear systems of equations for $U_1({\bm r},t)$ and ${\bm v}_1({\bm r},t)$, and $U_2({\bm r},t)$ and ${\bm v}_2({\bm r},t)$ read
\begin{equation} \label{eq:first3_main}
\left\{
\begin{array}{l}
    \left[\omega_c^2-\omega^2 f_\omega^2 \right] U_1({\bm r})-s^2f_\omega\nabla^2 U_1({\bm r})=0
    \vspace{0.2cm}\\
    {\displaystyle U_1({\bm r})\Big|_{\rm source}=U_{\rm ext}({\bm r}) }
    \vspace{0.2cm}\\
    {\hat {\bm n}}\cdot\big[ i\omega f_\omega\nabla U_1({\bm r})+\omega_c \hat{{\bm z}}\times\nabla U_1({\bm r})\big]\Big|_{\rm drain}=0
\end{array}
\right.
\end{equation}
where $s=\sqrt{-eU_0/m}$ is the plasma wave velocity, $f_\omega=1+i/(\omega\tau)$, and
\begin{equation} \label{eq:finalset_main}
\left\{
    \begin{array}{l}
    {\displaystyle \frac{1+(\tau\omega_c)^2}{U_0\tau}\nabla\cdot[U_1^*({\bm r}){\bm v}_1({\bm r})+U_1({\bm r}){\bm v}_1^*({\bm r})]=\nabla^2\phi({\bm r})}
    \vspace{0.2cm}\\
    {\displaystyle \phi({\bm r})-{\bm v}_1^*({\bm r})\cdot{\bm v}_1({\bm r})\Big|_{\rm source}=0}
    \vspace{0.2cm}\\
    {\displaystyle {\hat {\bm n}}\cdot\big[\omega_c\hat{{\bm z}}\times\nabla\phi({\bm r})-\frac{1}{\tau}\nabla\phi({\bm r})\big] \Big|_{\rm drain}=0}
    \end{array}
    \right.
    .
\end{equation}
Here, $\phi({\bm r})={\bm v}_1^*({\bm r})\cdot{\bm v}_1({\bm r})-eU_2({\bm r})/m$ and
\begin{equation}\label{eq:velocity_main}
    {\bm v}_1({\bm r})=\frac{s^2}{U_0}\frac{i\omega f_\omega \nabla U_1({\bm r})+\omega_c\hat{{\bm z}}\times\nabla U_1({\bm r})}{\omega_c^2-\omega^2f_\omega^2}
    ~.
\end{equation}
The Poisson problem in Eqs.~(\ref{eq:finalset_main}) admits a unique solution for $\phi({\bm r})$ and therefore for $U_2({\bm r}) = m/e\big[{\bm v}_1^*({\bm r})\cdot{\bm v}_1({\bm r}) - \phi({\bm r})\big]$. In the absence of a magnetic field, the photoresponse of the system will exhibit resonances at given frequencies dependant on the geometry of the system. The lowest of these frequencies is denoted as $\omega_B$ and determined numerically for any given disk geometry for later use (see the following sections). 

%
\begin{figure}[!t]
\begin{center}
\begin{tabular}{c}
\begin{overpic}[width=0.90\columnwidth]{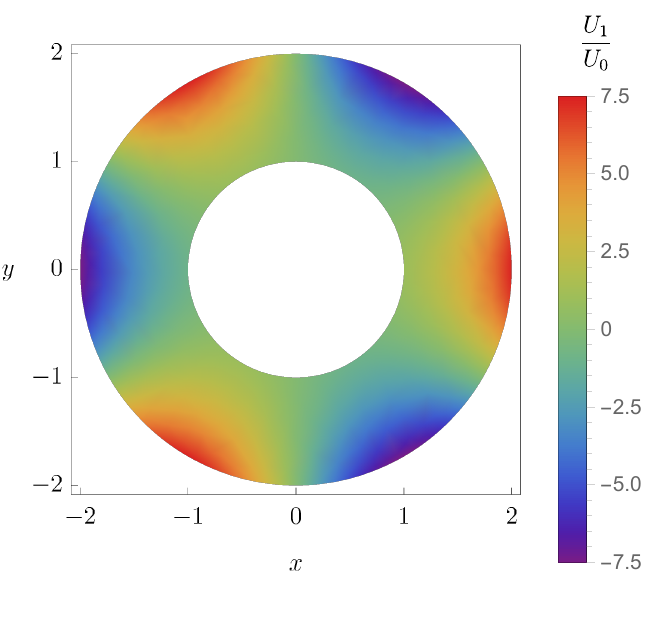}
\put(0,0){(a)}
\end{overpic}
\\
\begin{overpic}[width=0.90\columnwidth]{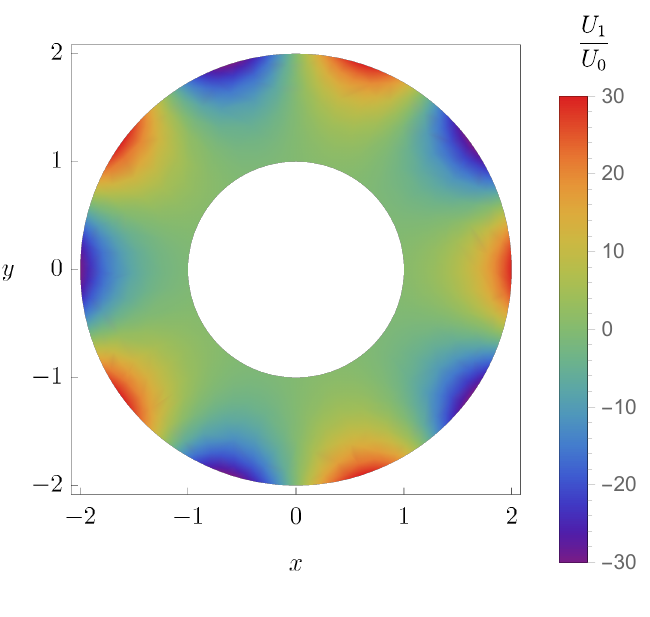}
\put(0,0){(b)}
\end{overpic}
\end{tabular}
\end{center}
    \caption{  Density plots of the real part of the linear potential, $U_1({\bm r})$ for $\omega=\omega_c=5\  s/r_0$. Panel (a) is evaluated at $\eta=3$. We can observe three complete oscillations of the potential around the circumference.
    Panel (b) is evaluated at $\eta=5$. Similarly, here we observe five complete oscillations. These are evaluated at $\tau=5\ r_0/s $, where $r_1=2\ r_0$.
    }
    \label{fig:edgeplasmonCorbino}
\end{figure}

\section{Concentric Corbino disk}
\label{sect:corbino_disk_analytic}
We first solve Eqs.~(\ref{eq:first3_main})-(\ref{eq:finalset_main}) for a concentric Corbino disk akin to the one studied in Ref.[\onlinecite{bandurin-corbino}], whose inner (source) and outer (drain) radii are $r_0$ and $r_1$, respectively. In this geometry, we can readily solve Eqs.~(\ref{eq:first3_main}) analytically and determine the full spectrum of magnetoplasmon modes, owing to the inherent rotational symmetry of the system. Such symmetry enables the separation of radial and angular variables within the solution. 
We note that our study differs from that of Ref.[\onlinecite{bandurin-corbino}] in two respects. Firstly, we consider the role of the magnetic field in modifying the spectrum of magnetoplasmons. Secondly, we consider the impact of source-to-gate voltages having a finite (integer) winding number $\eta$. We therefore impose that $U_1({\bm r})$ is equal to ${\bar U}_{\rm ext} \cos(\eta\theta)$ at the source, where ${\bar U}_{\rm ext}$ is the magnitude of the external potential and $\theta$ is the angle between ${\bm r}$ and the ${\hat {\bm x}}$-axis. Hence the linear solution will have winding numbers $\pm \eta$. Defining $k^2=(\omega_c^2-\omega^2 f_\omega^2){\rm sign}(\omega_c^2-\omega^2)/(s^2 f_\omega)$, the solutions of the system of linear differential equations~(\ref{eq:first3_main}) takes the form $U_1({\bm r}) = U_1^{(\eta)}(r,\theta) + U_1^{(-\eta)}(r,\theta)$, where
\begin{equation}\label{eq:bessel_solutions}
U_1^{(\eta)}(r,\theta) = \left\{
\begin{array}{ll}
\big[A_\eta I_\eta(kr)+B_\eta K_\eta(kr)]e^{i\eta\theta}~, & {\rm if}~ \omega^2<\omega_c^2
\vspace{0.2cm}\\
\big[C_\eta J_\eta(kr)+D_\eta Y_\eta(kr)]e^{i\eta\theta}~, & {\rm if}~ \omega^2>\omega_c^2
\end{array}
\right.
\end{equation}
Here, $J_\eta(x) = J_{-\eta}(x)$ [$I_\eta(x) = I_{-\eta}(x)$] and $Y_\eta(x) = Y_{-\eta}(x)$ [$K_\eta(x) = K_{-\eta}(x)$] are (modified) Bessel functions of the first and second kind, respectively.
The coefficients $A_\eta, B_\eta, C_\eta$ and $D_\eta$ are determined by applying the boundary conditions. 
After some lengthy but straightforward algebra we find, for $|\omega|<|\omega_c|$,
\begin{eqnarray}
    U_1^{(\eta)}(r,\theta)&=&\frac{{\bar U}_{ext}}{2}\Bigg[\frac{I_\eta(kr)}{I_\eta(kr_0)}-\frac{I'_\eta(kr_1)-\gamma_\eta I_\eta(kr_1)}{{\cal D}_\eta(\omega) I_\eta(kr_0)}
    \nonumber\\
    &\times&
    \left(\frac{K_\eta(kr)}{K_\eta(kr_0)}-\frac{I_\eta(kr)}{I_\eta(kr_0)}\right)\Bigg]e^{i\theta \eta},
    \label{eq:solutionfull} 
\end{eqnarray}
where $\gamma_\eta=\omega_c \eta/(\omega f_\omega k r_1)$, $I'_\eta(x)=dI_\eta(x)/dx$, $K'_\eta(x)=dK_\eta(x)/dx$, and
\begin{equation} \label{eq:relation3.2} 
    {\cal D}_\eta(\omega)=\frac{K'_\eta(kr_1)-\gamma_\eta K_\eta(kr_1)}{K_\eta(kr_0)}-\frac{I'_\eta(kr_1)-\gamma_\eta I_\eta(kr_1)}{I_\eta(kr_0)}.
\end{equation}
For $|\omega|>|\omega_c|$, $U_1^{(\eta)}(r,\theta)$ has the same form of Eqs.~(\ref{eq:solutionfull})-(\ref{eq:relation3.2}), with $J_n(k r)$ and $Y_n(k r)$ in lieu of $I_n(kr)$ and $K_n(kr)$, respectively. In Fig.~\ref{fig:edgeplasmonCorbino}~we plot the real part of the linear potential $U_1^{(\eta)}({\bm r})$. Counting oscillations at the outer perimeter of the disk (the drain), it can be seen that the two edge plasmons produced by manual injection at the source have $\eta=3$ [panel (a)] and $\eta=5$ [panel (b)], respectively.
In this figure we scale the electrical potential with $U_0$, lengths with the source radius $r_0$ and times with $r_0/s$.

\begin{figure}[t]
\begin{center}
\begin{tabular}{c}
\begin{overpic}[width=0.97\columnwidth]{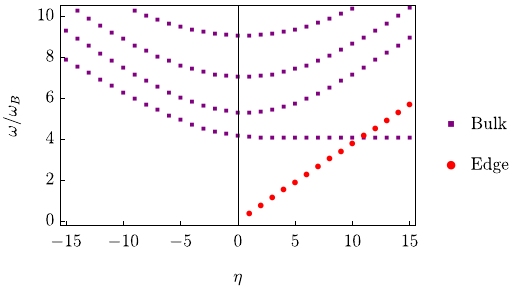}
\put(0,0){(a)}
\end{overpic}
\\
\begin{overpic}[width=0.97\columnwidth]{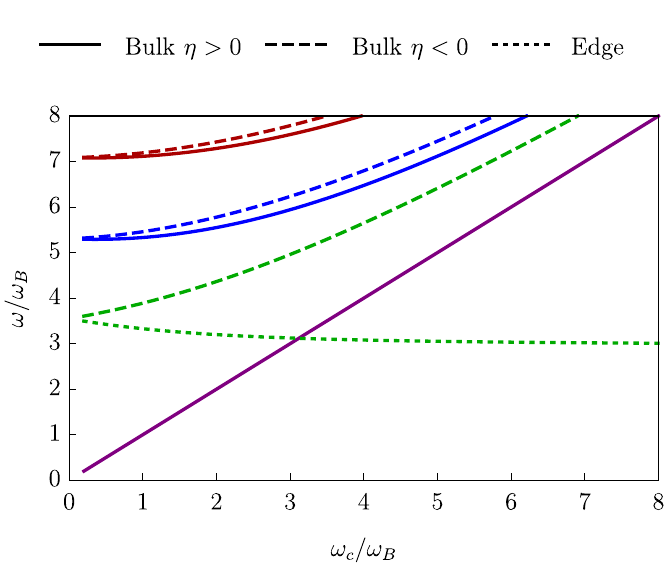}
\put(0,0){(b)}
\end{overpic}
\end{tabular}
\end{center}
    \caption{Panel (a) The resonant frequencies of the {\it linear} potential, $U_1^{(\eta)}(r,\theta)$, obtained from Eq.~(\ref{eq:solutionfull}) and plotted against the winding number $\eta$ defined before Eq.~(\ref{eq:bessel_solutions}). Bulk magnetoplasmon modes are represented by purple squares, while edge magnetoplasmons are represented by red circles. Panel (b) The resonant frequencies of the {\it linear} potential, $U_1^{(\eta)}({\bm r})$, against cyclotron frequency, at fixed $|\eta|=8$. Solid (dashed) lines refer to plasmons propagating in the counterclockwise (clockwise) direction. The dotted line denotes the edge state. The  purple solid line corresponds to the mode oscillating at the cyclotron frequency. Units are the same as in Fig.~\ref{fig:edgeplasmonCorbino}. Both figures are obtained in the limit $\tau\rightarrow\infty$.
    }
    \label{fig:modes}
\end{figure}

Bulk and edge magnetoplasmons can be identified as the zeros of ${\cal D}_\eta(\omega)$ and its counterpart for $|\omega|>|\omega_c|$. For $\omega_c>0$, the frequencies of magnetoplasmon modes as a function of the winding number $\eta$ are shown in Fig.~\ref{fig:modes}~(a).
In this figure potential, lengths and times are given in the same units of Fig.~\ref{fig:edgeplasmonCorbino}. For convenience, frequencies are scaled with the first resonant frequency at zero magnetic field, $\omega_B$.
There, ungapped edge modes are seen to wind in the $+{\hat{\bm \theta}}$ direction (as they only exist for positive $\eta$) and are localised at the outer edge of the disk. Winding in the opposite direction cannot occur as plasmons would be bound to the inner edge, which is however held at a fixed potential.

We also observe that bulk modes exhibit a variable degree of asymmetry: in general, the frequencies are higher for magnetoplasmons characterized by negative winding numbers. The asymmetry can be traced back to $\gamma_\eta$ defined after Eq.~(\ref{eq:solutionfull}), the only parameter that depends on the sign of $\eta$. Physically, this asymmetry arises from the relative alignment between the Lorentz force induced by the magnetic field, acting on the plasmons' constituent electrons, and the plasmons' electric field. The splitting in frequency of bulk modes can be observed in Fig.~\ref{fig:modes}~(b) where upper branches refer to negative winding numbers. In passing, we note that analogous splittings of frequencies of bulk modes have previously been observed in conventional disk geometries [\onlinecite{glattli,fetter2}] as well as Corbino disks with symmetric boundary conditions [\onlinecite{reboredo}]. In contrast, the present case, characterized by asymmetric boundary conditions, admits an additional nearly-flat band of normal modes. In fact, while the lowest branch of negative-$\eta$ bulk modes displays an approximately linear dispersion, positive-$\eta$ modes oscillate at around the cyclotron frequency. The latter plasmon nearly-flat band has no counterpart in conventional disks [\onlinecite{glattli,fetter2}] or Corbino disks under symmetric boundary conditions [\onlinecite{reboredo}]. We stress that the nearly-flat band becomes a clear feature of the spectrum only when the inner and outer radii of the Corbino disk are comparable. When this is not realized, it becomes unstable against the introduction of a small damping $1/\tau$, and the conventional disk solution is recovered [\onlinecite{glattli,fetter2}].

The flat plasmon band at $\omega \simeq \omega_c$ and $\eta>0$ in Fig.~\ref{fig:modes}~(a) has an important consequence for the nonlinear responsivity of the Corbino disk. For every external source-to-gate potential ${\bar U}_{\rm ext} \cos(\eta\theta)$, we expect the nonlinear rectified potential $U_2({\bm r})$ to exhibit a resonance at $\omega\simeq \omega_c$. In fact, $U_1(r_0,\theta)$ can be decomposed into the sum of two counter-winding potentials, characterized by winding numbers $\pm \eta$, one of which (depending on the direction of the magnetic field and the sign of $\omega_c$) can excite a magnetoplasmon mode at the cyclotron frequency. In turn, such mode produces a rectified voltage $U_2({\bm r})$ at the outer rim of the Corbino disk. We note that such voltage, thanks to the interference between oppositely-winding magnetoplasmons, not only is time-independent but it also contains a non-winding component characterized by $\eta=0$ that does not vanish when integrated over the drain. 

\begin{figure}[htp]
\begin{center}
\begin{tabular}{c}
\begin{overpic}[width=0.97\columnwidth]{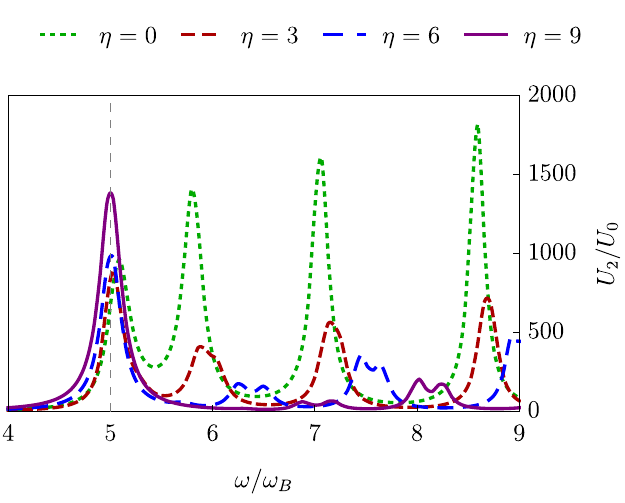}
\put(0,0){(a)}
\end{overpic}
\\
\begin{overpic}[width=0.97\columnwidth]{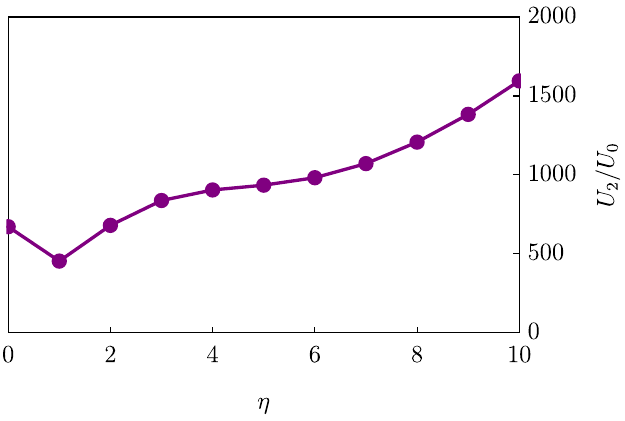}
\put(0,0){(b)}
\end{overpic}
\end{tabular}
\end{center}
    \caption{Panel (a) The {\it non-linear} potential at the drain, obtained by numerically solving Eqs.~(\ref{eq:finalset_main}), plotted against frequency, at $\omega_c=5\omega_B$. Different curves correspond to different winding numbers, $\eta$, of the source potential. We note that the first bulk mode remains pinned at $\omega=\omega_c$ and only increases in intensity with $|\eta|$, while all other modes slowly shift towards higher frequencies. The splitting of higher-order bulk modes becomes more and more evident at larger $\eta$: peaks split in two, as seen for e.g. $\eta=6$. Units are the same as in Figs.~\ref{fig:edgeplasmonCorbino} and~\ref{fig:modes}.
    For all curves we have set the collision time $\tau=5\ r_0/s$, and the outer radius, $r_1=2\ r_0$.
    Panel (b) The value of the nonlinear potential at $\omega=\omega_c$ as a function of $\eta$. The dip at $\eta=1$ is due to the fact that, for small values of $\eta$, the peak is slightly shifted to the right.
    }
    \label{fig:eta}
\end{figure}

In Fig.~\ref{fig:eta}(a) we show $U_2({\bm r})$, obtained by numerically solving Eq.~(\ref{eq:finalset_main}), integrated over the outer rim of the Corbino disk ({\it i.e.} the drain) for the first few values of $\eta$ and as a function of $\omega$. We clearly recognize a resonance at $\omega\simeq \omega_c$ for all values of $\eta$. In Fig.~\ref{fig:eta}(b), we show how the maximum of such resonance scales with $\eta$. 

Such result has an attractive implication. If we would be able to excite at once magnetoplasmons of frequency $\omega\simeq \omega_c$ in a broad range of winding numbers, the resulting resonance would grow to become particularly strong, therefore greatly enhancing the responsivity of the device. Furthermore, its position could be tuned by changing the external magnetic field, and it could be made to span the THz range practically at will. Unfortunately, the current geometry does not allow to easily achieve such result: to excite magnetoplasmons with different winding numbers it is necessary to carefully engineer the potential applied at the source. This requires superimposing various harmonics characterized by different values of $\eta$, a fact that is at present experimentally challenging.

For this reason, we will now move to study the experimentally more relevant case of an {\it eccentric} Corbino disk. In fact, while in the Corbino disk circular symmetry leads to the decoupling of various modes, the lack of symmetry of the eccentric disc allows their mixing. In turn, this enables the use of more realistic source potentials ({\it i.e.} uniform along the inner ring) to access the strong resonance at $\omega\simeq\omega_c$, as we proceed to show.


\section{Eccentric Corbino disk}
\label{sect:eccentric_corbino_disk}

\begin{figure}[t]
\begin{center}
\begin{overpic}[width=0.95\columnwidth]{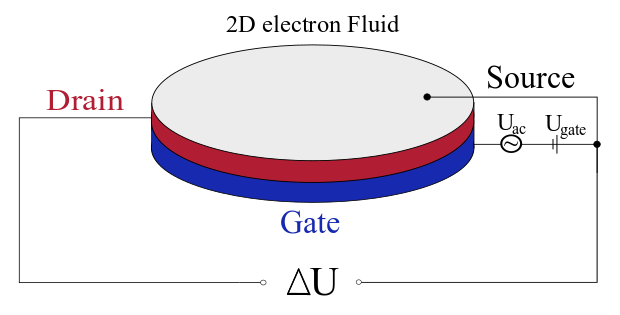}
\end{overpic}
\end{center}
    \caption{Schematic of the eccentric Corbino disk FET studied in this paper. The perimeter of the device acts as the drain while a finite small source is connected to the top of the cavity. $U_{\rm gate}$ is the back gate DC bias voltage, which in our case is constant in time and used to fix the charge density. The FET rectifies the AC source-gate voltage, $U_{\rm ac}(t)$, into the DC source-drain voltage $\Delta U$.
    }
    \label{fig:circuit}
\end{figure}
\begin{figure}[!htp]
\begin{center}
\begin{tabular}{c}
\begin{overpic}[width=0.95\columnwidth]{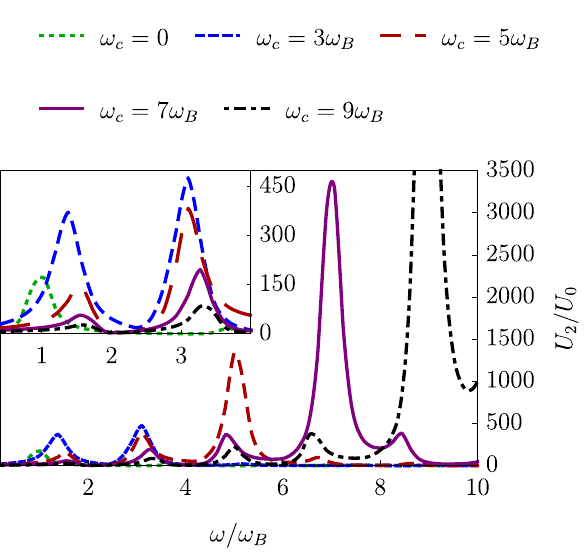}
\put(0,0){(a)}
\end{overpic}
\\
\begin{overpic}[width=0.95\columnwidth]{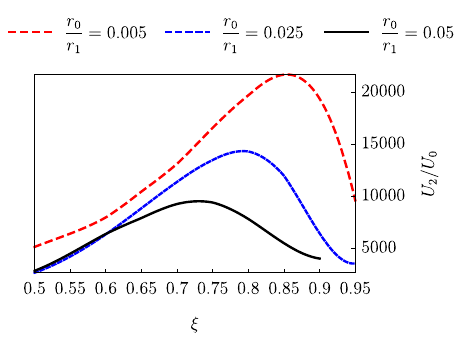}
\put(0,0){(b)}
\end{overpic}
\end{tabular}
\end{center}
    \caption{ 
    Panel (a) The nonlinear potential, $U_2({\bm r})$ integrated along the drain, obtained by numerically solving the set of Eqs.~(\ref{eq:first3_main})-(\ref{eq:finalset_main}), plotted as a function of the frequency of the incoming radiation. Different curves correspond to different values of the magnetic field, {\it i.e.} of the cyclotron frequency $\omega_c$. For all curves we have set the collision time $\tau=10\ r_1/2s$, and the inner radius $r_0=0.05\ r_1/2$, and thus the eccentricity is $\xi=d/r_1=0.95$. Additionally $U_{\rm ext}=U_0$.
    The inset shows a magnification of the graph for $0<\omega/\omega_B<4$. Here it is evident that the peak at $\omega\approx3\omega_B$ shifts to the right as $\omega_c$ increases.
    Panel (b) The strength of the peak of the nonlinear potential at the cyclotron frequency, for $\omega_c=7\omega_B$, plotted against eccentricity for different source radii. 
    The other parameters are the same of Panel (a).
    }
    \label{fig:peaks}
\end{figure}
The eccentric Corbino FET geometry is shown schematically in Fig.~\ref{fig:circuit}. In this geometry, the inner source ring is shrunk and placed off-centre.
The non-linear hydrodynamic problem, with the asymmetric boundary conditions of Eq.~(\ref{eq:boundary1}), can be solved numerically as described in Sect.~\ref{sect:cavity_model}. First, Eqs.~(\ref{eq:first3_main}) are solved for the linear potential. Then, by using Eq.~(\ref{eq:velocity_main}), Eqs.~(\ref{eq:finalset_main}) are solved for $\phi(\bm r)$. From the latter, we can then calculate the non-linear potential $U_2(\bm r)$. Since both Eqs.~(\ref{eq:first3_main}) and~(\ref{eq:finalset_main}) are Poisson problems, they admit unique solutions for a given set of boundary conditions.
We define the eccentricity as $\xi=d/r_1$ where $d$ is the distance of the centre of the source from the centre of the disk, and $r_1$ is the outer radius of the disk.
To aid comparison with the previous section, we keep the drain radius identical to that of the concentric Corbino disk, and we therefore scale lengths [times] with $r_1/2$ [$r_1/(2s)$]. Similarly, as in the previous section, frequencies will be scaled by $\omega_B$, the lowest resonance frequency at zero magnetic field determined numerically for any given geometry.

We plot the non-linear potential $U_2(\bm r)$, integrated along the drain, as a function of the AC driving frequency for various magnetic field strengths in Fig.~\ref{fig:peaks}~(a). For each curve, resonances at $\omega<\omega_c$ correspond to edge modes, while those at $\omega\geq \omega_c$ can be due to both bulk or edge ones. Now that the source has been placed off-centre and close to the drain, we can see that edge plasmons with differing winding numbers, and hence different frequencies, can propagate. As an example, for $\omega_c=7\omega_B$, we can see three edge modes below the cyclotron frequency (of frequencies $\omega/\omega_B\approx 1.5,\ 3.2,\ 4.8$) and one mode above it (at $\omega/\omega_B\approx 8.4$). As expected from the discussion in the previous section, for all field strengths the first bulk mode, fixed around the cyclotron frequency, results in the largest resonance peak.

It should be noted that although Fig.~\ref{fig:peaks}~(a) is obtained by setting the eccentricity $\xi=0.95$, this is not the optimum value that maximizes the photoresponse. In fact Fig.~\ref{fig:peaks}~(b) shows that for a source of radius $r_0=0.05\ r_1/2$, with $\omega_c=7\omega_B$ as used in panel~(a), the optimum eccentricity is $\xi\approx0.8$. Fig.~\ref{fig:peaks}~(b) further shows that the optimum eccentricity is inversely proportional to the source radius, $r_0$. It can be further shown that it increases with the drain radius, $r_1$, and cyclotron frequency, $\omega_c$. As such the geometry of such a device must be tailored to the expected frequency of incoming light.


We now wish to briefly comment on the feasibility of our device. We consider an FET based on doped bilayer graphene at relatively small ({\it i.e.} non-quantizing) magnetic fields, with dimensions on the order of a few micrometers: similar devices have been recently realised and shown to be significantly tunable via the application of gate voltage [\onlinecite{bandurin}]. Given the lowest bulk plasmon frequency of such devices [\onlinecite{bandurin,Vicarelli1,spirito,qin}], $\omega_B=300~{\rm GHz}$ (this is typically dependent on system size and for graphene can be changed via the gate voltage), and an effective electron mass [\onlinecite{bandurin}], $m\approx0.036m_e$, where $m_e$ is the free electron mass, we can estimate the lower limit for the magnetic field. The lowest observable edge plasmon frequency is always similar to the lowest bulk plasmon frequency provided the source radius is small, thus, by equating the lowest bulk plasmon frequency with the cyclotron frequency, $\omega_c=eB/m^*$, our estimate for the minimum magnetic field becomes $B_{\rm min}\approx0.06~{\rm T}$. This magnitude is easily achievable in experiments. In passing, we mention that alternatives to applying an external magnetic field do exist [\onlinecite{potashin,principi4}].

\section{Conclusions}
\label{sect:conclusions}
In this paper we have studied Corbino-disk-shaped photodetectors with sources at the inner ring which oscillate at the frequency of the incoming radiation with respect to metallic back-gates. The design is similar to that of conventional Dyakonov-Shur devices, in that a rectified potential is measured at the outer rim of the disk, which acts as a fluctuating drain. By applying a magnetic field in the direction perpendicular to the cavity, the rectification of long-wavelength radiation occurs from the constructive interference of not only bulk plasmons, but also edge magnetoplasmons.

In this geometry, plasmons can circulate along the entirety of the disk's perimeter nearly unimpeded [\onlinecite{ashoori}]. Plasmons in this configuration are categorised by their winding number, {\it i.e.} the number of complete oscillations of the electron density that occur over a full revolution around the disk. In the first part of the paper, we studied the response of a conventional Corbino-disk photodetector with the source-ring located at the centre of the disk. Said geometry admits an analytic solution. In this configuration individual plasmon modes can be manually injected by selecting the winding number of the external source-to-gate potential. It is important to note that, as shown in Sect.~\ref{sect:corbino_disk_analytic}, all  modes, and in particular ones at $\omega\simeq \omega_c$ (which exist only in the presence of asymmetric boundary conditions and up to large winding numbers, as long as inner and outer radii are comparable and damping is small), produce nonlinear rectified potentials that are also uniform along the edge. Therefore, all contributions at $\omega\simeq \omega_c$ can in principle be summed up, with a careful choice of the source-to-gate external potential, and result in a large resonance at the cyclotron frequency that greatly enhances the responsivity of the device. Since its frequency depends on the magnetic field, exploiting such strong resonance can lead to the realization of efficient and tunable THz photodetectors. Unfortunately, this programme is difficult to be achieved in practice. 

Instead, through breaking the circular symmetry of the system by placing the source off-center and closer to the edge of the disk, magnetoplasmons with various winding numbers can be excited with source-to-gate voltages easily achievable experimentally ({\it i.e.} uniform along the source perimeter).
By tuning the degree of eccentricity of the system, we are able to excite various magnetoplasmons at once. Therefore, we are able to enhance the photodetector responsivity at the frequency range corresponding to the cyclotron one. The best protocol for photodetection clearly depends on one's aims. When searching for the frequency of incoming radiation, it is best to fix the luminosity of the radiation, where possible, and scan over a presumed range of frequencies by changing the magnetic field strength. When measuring the luminosity of incoming radiation it is best to adjust the cyclotron frequency to match the incoming radiation's frequency to achieve a high gain. 

\acknowledgments
A.P. acknowledges support from the European Commission under the EU Horizon 2020 MSCA-RISE-2019 programme (project 873028 HYDROTRONICS) and of the Leverhulme Trust under
the grant RPG-2019-363.

\appendix

\section{Derivation of Eqs.~(\ref{eq:first3_main})-(\ref{eq:finalset_main})}
\label{app:derivation_model}
Plugging the expansions in Eq.~(\ref{eq:expansion1}) into the set of equations~(\ref{eq:local_gate_approx})-(\ref{eq:boundary1}), we collect terms of order $U_{\rm ext}$ and $U_{\rm ext}^2$ into two systems of {\it linear} differential equations, {\it i.e.}
\begin{equation} \label{eq:first}
\left\{
\begin{array}{l}
    \partial_t U_1({\bm r},t) =-U_0\nabla\cdot {\bm v}_1({\bm r},t)
    \vspace{0.2cm}\\
    {\displaystyle \frac{e}{m}\nabla U_1({\bm r},t) =\partial_t{\bm v}_1({\bm r},t)+ \frac{1}{\tau}{\bm v}_1({\bm r},t)+\omega_c \hat{{\bm z}}\times{\bm v}_1({\bm r},t) }
    \vspace{0.2cm}\\
    U_1({\bm r},t)|_{\rm source} =U_{\rm ext}\cos(\omega t)
    \vspace{0.2cm}\\
    {\hat {\bm n}}\cdot{\bm v}_1({\bm r},t)|_{\rm drain} =0
\end{array}
\right.
,
\end{equation}
and
\begin{equation} \label{eq:second}
\left\{
\begin{array}{l}
    \partial_t U_2({\bm r},t)= -\nabla\cdot [U_0{\bm v}_2({\bm r},t)+U_1({\bm r},t){\bm v}_1({\bm r},t)]
    \vspace{0.2cm}\\
    {\displaystyle -\nabla \phi({\bm r},t) =\partial_t{\bm v}_2({\bm r},t)+ \frac{1}{\tau}{\bm v}_2({\bm r},t)+\omega_c \hat{{\bm z}}\times{\bm v}_2({\bm r},t)}
    \vspace{0.2cm}\\
    U_2({\bm r},t)|_{\rm source}=0,
    \vspace{0.2cm}\\
    {\hat {\bm n}}\cdot{\bm v}_2({\bm r},t)|_{\rm drain}=0.
\end{array}
\right.
,
\end{equation}
respectively.
Here we defined $\phi({\bm r},t) = v_1^2({\bm r},t)/2-eU_2({\bm r},t)/m$. Eqs.~(\ref{eq:first}) form a closed set of linear differential equations that can be solved exactly. Their result is then substituted into Eqs.~(\ref{eq:second}), which are themselves linear in $U_2({\bm r},t)$ and ${\bm v}_2({\bm r},t)$ and whose solution yields the rectified potential. The second order set of equations~(\ref{eq:second}) can be simplified further by noting that we are looking for a time-independent potential, therefore by integrating over a period of oscillation, $T=2\pi/\omega$, the time-dependent parts of $U_2({\bm r},t)$ and ${\bm v}_2({\bm r},t)$ will vanish. For a generic function of time $A(t)$, we define its time-average as
\begin{equation}
    \braket{A(t)}=\frac{1}{T}\int_0^T A(t)dt
    ~.
\end{equation}
After time averaging, Eq.~(\ref{eq:second}) becomes
\begin{equation} \label{eq:second2}
\left\{
\begin{array}{l}
    \nabla\cdot [U_0{\bm v}_2({\bm r})+\braket{U_1({\bm r},t){\bm v}_1({\bm r},t)}]=0
    \vspace{0.2cm}\\
    {\displaystyle \frac{1}{\tau}{\bm v}_2({\bm r})+\omega_c \hat{{\bm z}}\times{\bm v}_2({\bm r}) = - \nabla \phi({\bm r})}
    \vspace{0.2cm}\\
    U_2({\bm r})|_{\rm source}=0
    \vspace{0.2cm}\\
    {\hat {\bm n}}\cdot{\bm v}_2({\bm r})|_{\rm drain}=0
\end{array}
\right.,
\end{equation}
where now $\phi({\bm r},t) = \braket{v_1^2({\bm r},t)}/2-eU_2({\bm r})/m$, and $U_2({\bm r})$ and ${\bm v}_2({\bm r})$ denote the time-independent components of $U_2({\bm r},t)$ and ${\bm v}_2({\bm r},t)$, respectively. 

We will now further simplify Eqs.~(\ref{eq:first}). We first obtain two equations by applying the operator $\partial_t +1/\tau$ and the cross product with ${\hat {\bm z}}$ to the second of Eqs.~(\ref{eq:first}). We then combine the two equations we obtained, and get
\begin{eqnarray} \label{eq:cartesian3}
    \left[\left(\partial_t+\frac{1}{\tau}\right)^2+\omega_c^2 \right]{\bm v}_1({\bm r},t)&=&\frac{e}{m}\Bigg[\left(\partial_t+\frac{1}{\tau}\right)\nabla U_1({\bm r},t)
    \nonumber\\
    &-&\omega_c \hat{{\bm z}}\times\nabla U_1({\bm r},t)\Bigg].
\end{eqnarray}
The new set of equations is solved by using the {\it Ansatz} (see also the main text, Sect.~\ref{sect:cavity_model})
\begin{equation}
    \begin{aligned}
    &U_1({\bm r},t)=U_1({\bm r})e^{-i\omega t}+U_1^*({\bm r})e^{i\omega t},\\
    &{\bm v}_1({\bm r},t)={\bm v}_1({\bm r})e^{-i\omega t}+{\bm v}_1^*({\bm r})e^{i\omega t},
    \label{eq:trial}
    \end{aligned}
\end{equation}
from which we obtain the following set of time-independent linear equations:
\begin{equation} \label{eq:first2_1}
-i\omega U_1({\bm r})+U_0\nabla\cdot {\bm v}_1({\bm r})=0~,
\end{equation}
and
\begin{equation} \label{eq:first2_2}
[\omega^2 f_\omega^2 - \omega_c^2]{\bm v}_1({\bm r})=\frac{e}{m}\big[i\omega f_\omega\nabla U_1({\bm r})
+\omega_c \hat{{\bm z}}\times\nabla U_1({\bm r})\big]
~,
\end{equation}
subject to the boundary conditions
\begin{equation} \label{eq:first2_3}
    \begin{aligned}
    &U_1({\bm r})|_{\rm source}=\frac{U_{\rm ext}}{2},\\
    &{\hat {\bm n}}\cdot{\bm v}_1({\bm r})|_{\rm drain}=0.
    \end{aligned}
\end{equation}
In these equations we introduced 
$f_\omega=1+i/(\omega\tau)$. 
In addition to Eqs.~(\ref{eq:first2_1})-(\ref{eq:first2_3}), we have a set of equation for the quantities $U_1^*({\bm r})$ and ${\bm v}_1^*({\bm r})$. These are obtained from Eqs.~(\ref{eq:first2_1})-(\ref{eq:first2_3}) by taking their complex conjugates. Substituting Eq.~(\ref{eq:first2_2}) into~(\ref{eq:first2_1}), results in the following closed set of equations for $U_1({\bm r})$:
\begin{equation} \label{eq:first3}
\left\{
\begin{array}{l}
    \left[\omega_c^2-\omega^2 f_\omega^2 \right] U_1({\bm r})-s^2f_\omega\nabla^2 U_1({\bm r})=0
    \vspace{0.2cm}\\
    {\displaystyle U_1({\bm r})\Big|_{\rm source}=\frac{U_{\rm ext}}{2}}
    \vspace{0.2cm}\\
    {\hat {\bm n}}\cdot\big[ i\omega f_\omega\nabla U_1({\bm r})+\omega_c \hat{{\bm z}}\times\nabla U_1({\bm r})\big]\Big|_{\rm drain}=0
\end{array}
\right.
\end{equation}
Here we define the plasma wave velocity, $s=\sqrt{-eU_0/m}$, where $U_0$, the equilibrium potential, is negative for an electron fluid. The first of Eqs.~(\ref{eq:first3}) defines a Poisson problem which, once boundary conditions are specified as in the second and third of~(\ref{eq:first3}), admits a unique solution. Such solution is determined analytically for the case of a concentric Corbino-disk geometry in Sect.~\ref{sect:corbino_disk_analytic} and numerically for an eccentric disk in Sect.~\ref{sect:eccentric_corbino_disk}. 

Once the set of Eqs.~(\ref{eq:first3}) is solved and $U_1({\bm r})$ has been determined, the velocity is given by
\begin{equation}\label{eq:velocity}
    {\bm v}_1({\bm r})=\frac{s^2}{U_0}\frac{i\omega f_\omega \nabla U_1({\bm r})+\omega_c\hat{{\bm z}}\times\nabla U_1({\bm r})}{\omega_c^2-\omega^2f_\omega^2}
    ~.
\end{equation}
It is then possible to approach the problem posed by the set of Eqs.~(\ref{eq:second2}) in a similar fashion. Plugging the definitions in Eqs.~(\ref{eq:trial}) in there, we find
\begin{equation} \label{eq:second3}
    \left\{
    \begin{array}{l}
    \nabla\cdot [U_0{\bm v}_2({\bm r})+U_1^*({\bm r}){\bm v}_1({\bm r})+U_1({\bm r}){\bm v}_1^*({\bm r})]=0
    \vspace{0.2cm}\\
    {\displaystyle \frac{1}{\tau}{\bm v}_2({\bm r})+\omega_c \hat{{\bm z}}\times{\bm v}_2({\bm r})= - \nabla\phi({\bm r}) }
    \vspace{0.2cm}\\
    U_2({\bm r})|_{\rm source}=0
    \vspace{0.2cm}\\
    {\bm v}_2({\bm r})|_{\rm drain}=0
    \end{array}
    \right.
\end{equation}
where, explicitly, $\phi({\bm r})={\bm v}_1^*({\bm r})\cdot{\bm v}_1({\bm r})-eU_2({\bm r})/m$. 
To further simplify Eq.~(\ref{eq:second3}) and reduce it to a Poisson problem, we first obtain two equations by taking the divergence and applying the operator ${\hat {\bm z}}\times \nabla$ to the second of its equations. We get
\begin{equation} \label{eq:second4_1}
    \frac{1}{\tau}\nabla\cdot{\bm v}_2({\bm r})-\omega_c \hat{{\bm z}}\cdot\nabla\times{\bm v}_2({\bm r})=\nabla^2\phi({\bm r})
    ~,
\end{equation}
and
\begin{equation} \label{eq:second4_2}
    \frac{1}{\tau}\hat{{\bm z}}\cdot\nabla\times{\bm v}_2({\bm r})+\omega_c\nabla\cdot{\bm v}_2({\bm r})=0
    ~.
\end{equation}
Combining such equations with the first of Eqs.~(\ref{eq:second3}) gives
\begin{equation}\label{eq:second5}
    \frac{1+(\tau\omega_c)^2}{U_0\tau}\nabla\cdot[U_1^*({\bm r}){\bm v}_1({\bm r})+U_1({\bm r}){\bm v}_1^*({\bm r})]=\nabla^2\phi({\bm r})
    ~.
\end{equation}
Eq.~(\ref{eq:second5}) has the form of a Poisson equation for $\phi({\bm r})$. Given appropriate boundary conditions, the latter can be solved and yield a unique solution for $\phi({\bm r})$ and therefore for $U_2({\bm r}) = m/e\big[{\bm v}_1^*({\bm r})\cdot{\bm v}_1({\bm r}) - \phi({\bm r})\big]$. To determine the boundary conditions for $\phi({\bm r})$, we first take the cross product of the second of Eqs.~(\ref{eq:second3}) with $\hat{{\bm z}}$, which yields
\begin{equation} \label{eq:second6}
    \frac{1}{\tau}\hat{{\bm z}}\times{\bm v}_2({\bm r}) -\omega_c {\bm v}_2({\bm r})=-\hat{{\bm z}}\times\nabla\phi({\bm r})
    ~.
\end{equation}
Substituting this back into the second of Eqs.~(\ref{eq:second3}) we get
\begin{equation}
    \big[1+(\omega_c\tau)^2\big]{\bm v}_2({\bm r})=\omega_c\tau^2 \hat{{\bm z}}\times\nabla\phi({\bm r})-\tau\nabla\phi({\bm r})
    ~.
\end{equation}
This leads us to the following solvable set of differential equations in $\phi({\bm r})$:
\begin{equation} \label{eq:finalset}
\left\{
    \begin{array}{l}
    {\displaystyle \frac{1+(\tau\omega_c)^2}{U_0\tau}\nabla\cdot[U_1^*({\bm r}){\bm v}_1({\bm r})+U_1({\bm r}){\bm v}_1^*({\bm r})]=\nabla^2\phi({\bm r})}
    \vspace{0.2cm}\\
    {\displaystyle \phi({\bm r})-{\bm v}_1^*({\bm r})\cdot{\bm v}_1({\bm r})\Big|_{\rm source}=0}
    \vspace{0.2cm}\\
    {\displaystyle \omega_c\hat{{\bm z}}\times\nabla\phi({\bm r})-\frac{1}{\tau}\nabla\phi({\bm r})\Big|_{\rm drain}=0}
    \end{array}
    \right.
    .
\end{equation}

\noindent \bibliographystyle{apsrev4-1}
\bibliography{main}

\end{document}